\documentclass{PoS}

\usepackage{amssymb,graphicx,amsmath, amsthm,verbatim,adjustbox}

\title{The Dark Matter Programme of the Cherenkov Telescope Array
}

\ShortTitle{The Dark Matter Programme of the Cherenkov Telescope Array}

\author{The CTA Consortium\thanks{ see http://www.cta-observatory.org/consortium\_authors/authors\_2017\_07.html for full author list } ~ {\normalfont represented by } \speaker{A.Morselli}$^1$\thanks{aldo.morselli@roma2.infn.it} \\
$^1$      INFN Roma Tor Vergata, Roma, Italy \\
}

\abstract{In the last decades a vaste amount of evidence for the existence of dark matter has been accumulated. At the same time, many efforts have been undertaken to try to identify what dark matter is. Indirect searches look at places in the Universe where dark matter is believed to be abundant and seek for possible annihilation or decay signatures. The Cherenkov Telescope Array (CTA) represents the next generation of imaging Cherenkov telescopes and, with one site in the Southern hemisphere and one in the Northern hemisphere, will be able to observe all the sky with unprecedented sensitivity and angular resolution above a few tens of GeV. The CTA Consortium will undertake an ambitious program of indirect dark matter searches for which we report here the brightest prospects.}

\FullConference{35th International Cosmic Ray Conference – ICRC217-\\
		10-20 July, 2017\\
		Bexco, Busan, Korea}

\begin{document}

\section{Introduction}

The existence of dark matter (DM) in our Universe  is well established, but its nature is at present still unknown. Multiple hypotheses endure as to the character of dark matter and for the most popular models discussed CTA has a unique chance of discovery. Among the most promising particle candidates are Weakly Interacting Massive Particles (WIMPs), which typically can self-annihilate to produce prompt or secondary gamma-rays during the annihilation. If WIMPs are produced thermally in the early Universe then the current self-annihilation cross-section has a natural value of approximately 3 $\cdot$10$^{-26} cm{^3}s^{-1}$ \cite{Steigman}. WIMPs models, such as the supersymmetric neutralino, give predictions for gamma-ray energy spectra from the annihilations, which are crucial inputs, together with the DM distribution in the observed target, to estimate prospects for the sensitivity of indirect searches.
The goal of the present study is to provide preliminary comparative expectations on indirect DM searches with the Cherenkov Telescope Array (CTA) \cite{CTA13, Doro, CTA15, CTA17} taking into account continuum gamma-ray signatures coming from typical DM annihilation channels.

The expected DM annihilation gamma-ray flux from a DM-dominated region depends  on the so-called particle physics and astrophysical (or $J$) factors: 

\begin{equation}
\Phi_{s}(\Delta\Omega)=\frac{1}{4\pi}\frac{\langle \sigma v \rangle}{2m^{2}_{DM}}\int^{E_{max}}_{E_{min}}\frac{dN_{\gamma}}{dE_{\gamma}}dE_{\gamma} \times J(\Delta\Omega),
\end{equation}
where $\langle \sigma v \rangle$ is the velocity-averaged self-annihilation cross-section, $m_{DM}$ is the dark matter particle mass, $E_{min}$ and $E_{max}$ are the energy limits for the measurement and $\frac{dN_{\gamma}}{dE_{\gamma}}$ is the energy spectrum of the gammas produced in the annihilation (as, e.g., from \cite{Cirelli}). The J-factor is the integral along the line of sight of the squared DM density profile of the given target integrated within an aperture angle, $\int_{\Delta\Omega}\text{d}\Omega\int_{l.o.s.}\rho^{2}_{DM}(\boldmath{r})\text{d}l$. The products of DM annihilation are thought to come from decay and/or hadronization of the primary Standard Model (SM) particles: quark-antiquark, lepton and boson, and each channel is expected to have its own branching ratio. 

The priority for the CTA DM program is to discover the nature of DM. The publication of limits following non-detection would certainly happen, but in planning the observational strategy, the priority of discovery drives the programme. The possibility of discovery should be considered in the light of model predictions where the minimum goal for searches within the most widely considered models is the thermal cross-section and CTA can probe DM scenarios out of reach for collider searches and direct detection experiments \cite{CTADS}. The principal target for DM observations in CTA is the Galactic halo. This observation will be taken within several degrees of the Galactic centre with the Galactic centre and the most intense diffuse emission regions removed from the analysis. With a cuspy DM profile, observations of 500 hours in this region provide sensitivities below the thermal cross-section and give a significant chance of discovery in some of the most popular models for WIMPs. Since the DM density in the Galactic halo is far from certain, and considering the irreducible background from cosmic rays and the galactic diffuse emission there, other targets are also proposed for observation. Among these secondary targets, the first to be observed will be dwarf spheroidal satellite galaxies with 100 hours per year proposed.
Beyond these two observational targets, alternatives will be considered closer to the actual date of CTA operations. New sky surveys such as the Dark Energy Survey (DES) \cite{DES} will extend the knowledge of possible sites of large DM concentrations and a detailed study of the latest data will be made to continuously select the best targets for DM searches in CTA. Among these new possible targets are newly discovered candidate dwarf galaxies and not yet discovered DM clumps (pure DM overdensities without baryonic counterpart and thus deprived of any stellar activity) which could be very promising sources if their locations are identified a priori by their gravitational effects. Beyond the targets proposed for observations in the present Key Science Projects (KSP), the data from the Large Magellanic Cloud (LMC) and from the clusters of Galaxies KSPs will also be used to search for DM. Furthermore, the data from the Galactic plane and extragalactic surveys might give hints of gamma-ray sources which do not have counterparts in multi-wavelength data and which could be pursued as DM targets.  
The DM Programme is very well suited to being carried out in the Core Programme by the CTA Consortium. The observations require a large amount of time with a significant chance of major discovery but with a clear risk of a null detection. For the Galactic halo the observation time used will also be of great use for astrophysics. In the phase between now and the actual operation of CTA, much will evolve in the knowledge of DM distributions in the various targets. Further detailed work is needed to understand the systematics in the backgrounds especially in the Galactic halo that looks the most promising target up to now.

\section{Galactic halo}

\begin{figure*}[h!t]
\centering
\begin{tabular}{cc}
\includegraphics[width=0.49\linewidth]{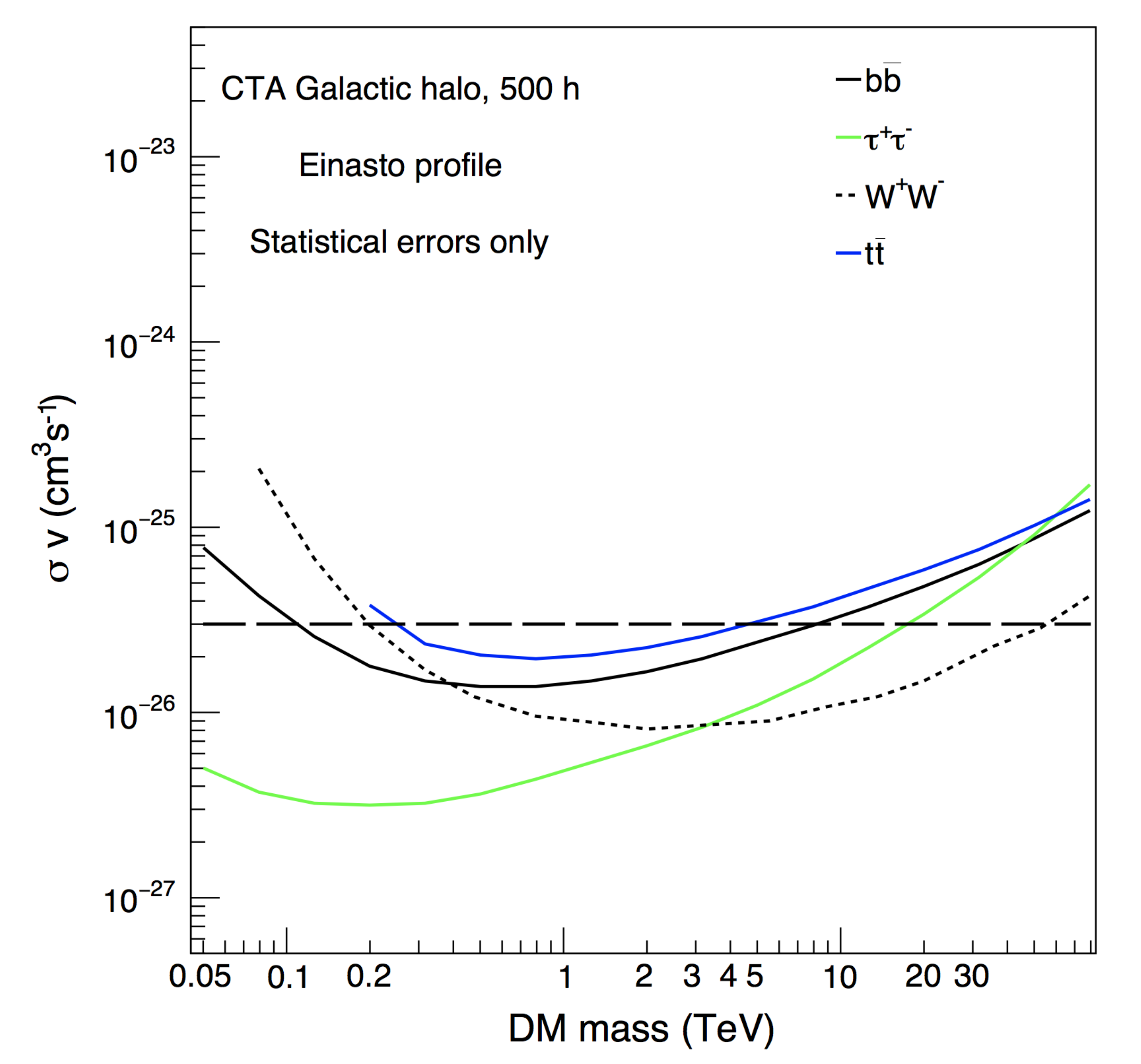}
\includegraphics[width=0.51\linewidth]{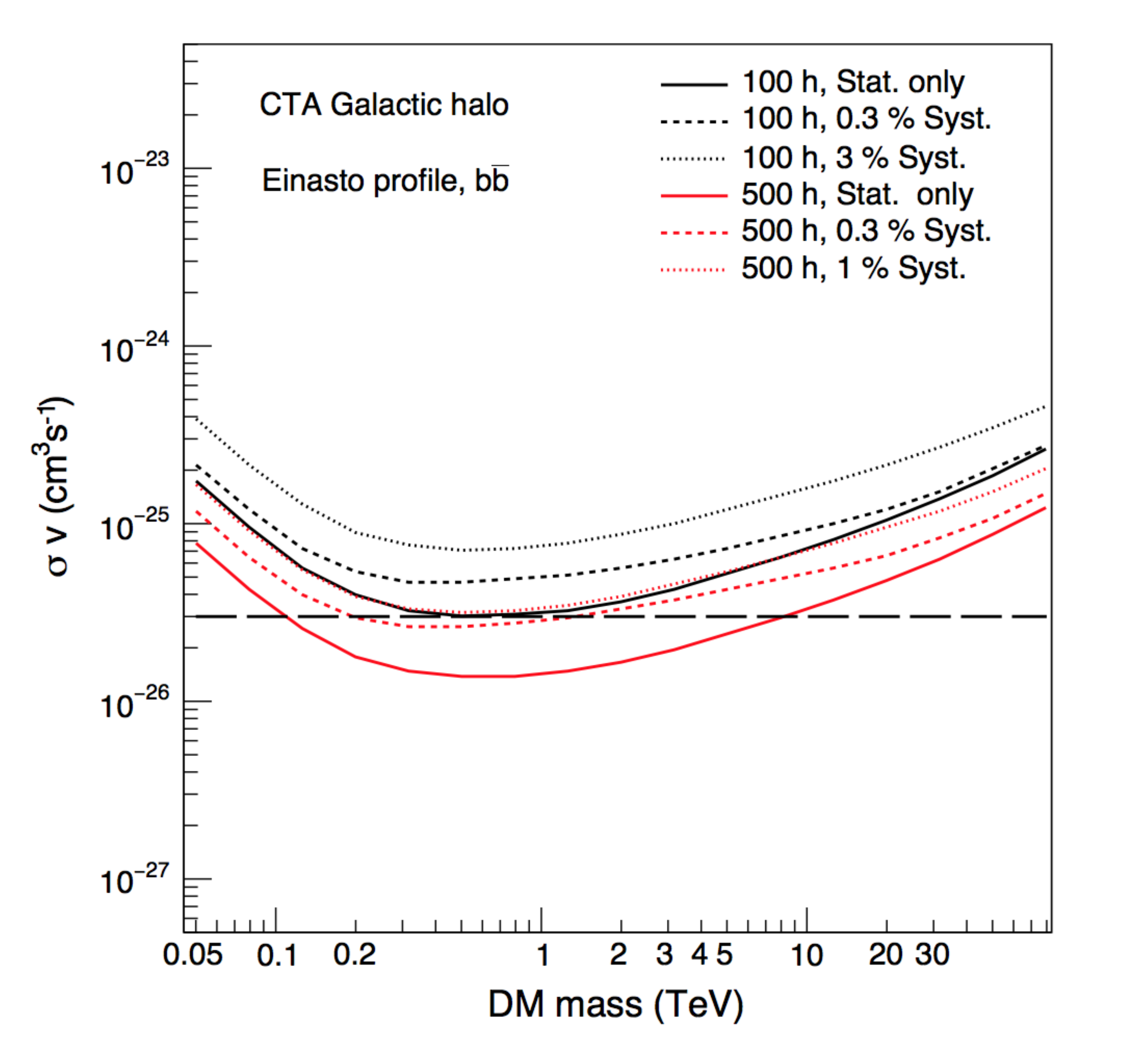}
\end{tabular}
\caption{Left panel: CTA sensitivity  from observation of the Galactic halo for the Einasto profile and for different annihilation modes as indicated. Only statistical errors were included. Right panel: CTA sensitivity for \textit{b\={b}} annihilation mode for the Einasto profile and for different conditions, black is for 100 hours of observation and red is for 500 hours. The solid lines are the sensitivities only taking into account the statistical errors while the dashed and dotted curves take into account systematics as indicated. The dashed horizontal lines approximate the level of the thermal relic cross-section.}
\label{fig:Halo}
\end{figure*}
\begin{figure}
\centering
\includegraphics[width=0.60\linewidth]{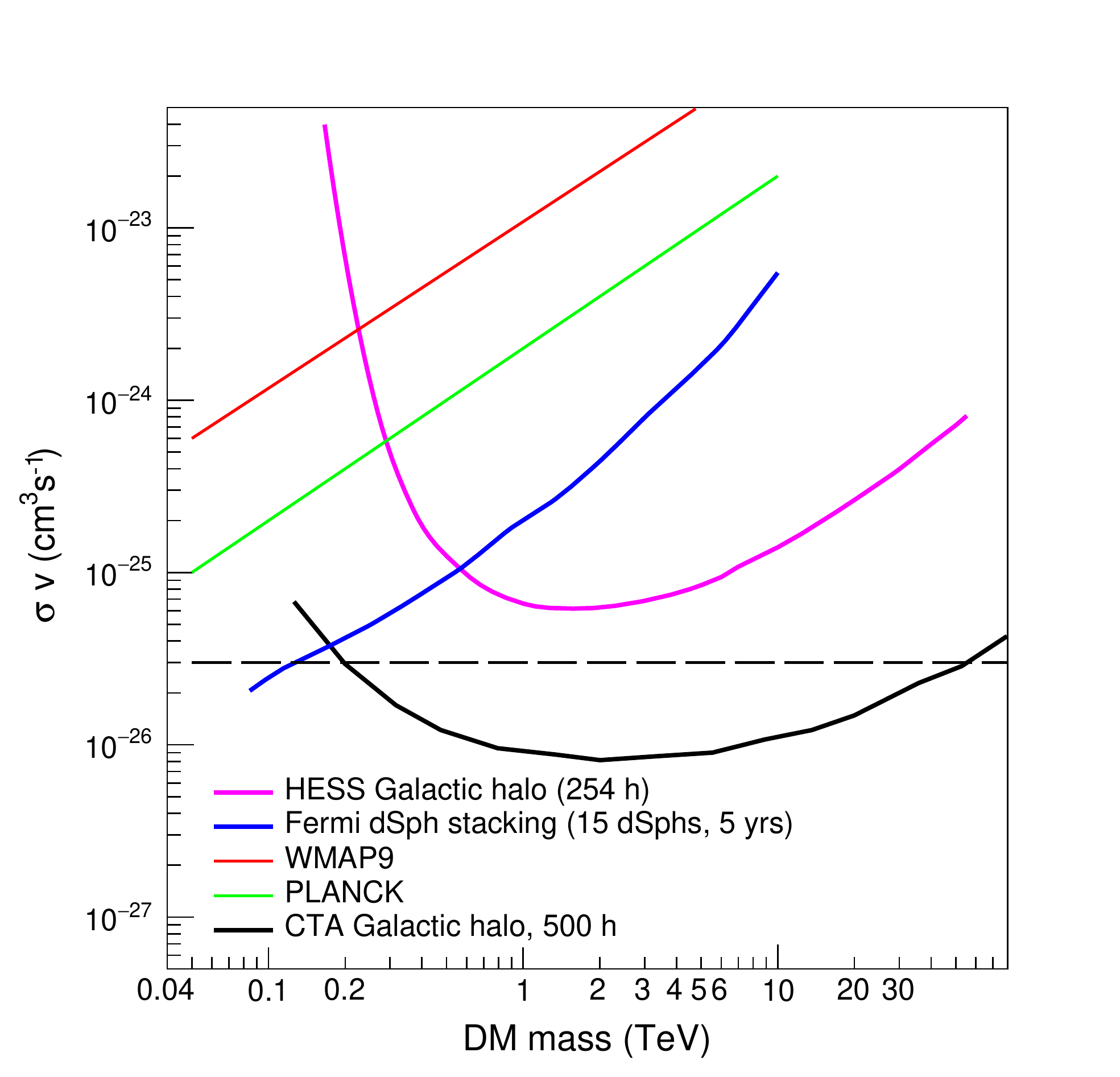}
\caption{Current best limits on the annihilation cross-section from indirect detection (Fermi-LAT dwarf spheroidal galaxies stacking analysis, $W^+W^-$ channel  \cite{Fermi_Dwarf},   H.E.S.S. Galactic halo $W^+W^-$ channel \cite{HESS2017}) and cosmic microwave background (WMAP and Planck  \textit{b\={b}} channel \cite{Plank}) experiments compared with the projected sensitivity for CTA from observations of the Galactic halo for the Einasto profile,  $W^+W^-$ channel.  The expectation for CTA is optimistic as it includes only statistical errors.}
\label{fig:halo_fig}
\end{figure}

The centre of the Milky Way has in the past been considered as a target for dark matter searches \cite{HESS2006, HESS2017}. 
More recently, because of the rich field of very high-energy (VHE) gamma-ray astrophysical sources in the region, the searches focus on the Galactic halo excluding the central region of Galactic latitude $b < 0.3^\circ$. Even excluding the very central region, the total mass of dark matter in the Galactic halo together with its proximity to Earth makes it the most promising source for dark matter searches with CTA. 
The inconvenience of this target, however, is the fact that being a diffuse source, the integration over the inner halo, while yielding a large signal, gives a very large instrumental background from misidentified charged cosmic rays \cite{Silverwood}.
 Furthermore, there are astrophysical backgrounds from various sources which must be understood, even with the very central region excluded from the analysis. It is believed that the disadvantages of the Galactic halo can be overcome with sufficient experimental effort to control systematic effects in the background subtraction and modeling. The expertise required for this analysis strengthens the case for this programme to be conducted by the CTA Consortium.
The Galactic halo observations will be taken with multiple grid pointings with offsets from the Galactic centre position of about $\pm 1.3^\circ$  to cover the central 4$^\circ$  as uniformly as possible. In the Galactic centre KSP a further 300 hours are proposed for astrophysics covering up to latitudes $\pm 10 ^\circ$. These data will also be included in the analysis for dark matter to improve the sensitivity for cored dark matter density profiles. Given the major scientific impact of a positive result, we propose that these observations are done in the first three years of CTA operation with high priority.
The sensitivity predictions for observations in the Galactic Halo are shown in Figure \ref{fig:Halo}.  The left-hand plot shows the sensitivity for different annihilation modes
 and the right-hand plot for different observation times with a method \cite{Lefranc} to include systematic uncertainties on the residual cosmic-ray background as indicated in the caption.

In Figure \ref{fig:halo_fig} is shown the CTA sensitivity to a WIMP annihilation signature as a function of WIMP mass for nominal parameters and for the multiple CTA observations. The dashed horizontal line indicates the likely cross-section for a WIMP which is a thermal relic of the Big Bang.

The predictions shown here can be considered optimistic, even when systematics errors are included, as we do not consider the effect of the Galactic diffuse emission as background for DM searches. This will be investigated in detail in a forthcoming publication by the CTA Consortium

\section{Dwarf satellite galaxies}

The dwarf spheroidal galaxies (dSphs) of the Local Group could give a clear and unambiguous detection of dark matter. They are gravitationally bound objects and are believed to contain up to $10^3$ times more mass in dark matter than in visible matter, making them widely discussed as potential targets. Being small and distant, many of the dwarf galaxies will appear as near point sources in CTA and hence the nuisance of the instrumental background is much reduced. Although less massive than the Milky Way or the LMC, they are also environments with a favourably low astrophysical gamma-ray background making the unambiguous identification of a dark matter signal easier compared to the Galactic centre or LMC. Neither astrophysical gamma-ray sources (supernova remnants, pulsar wind nebulae,...) nor gas acting as target material for cosmic rays have been observed in these systems.

\begin{figure*}[h!t]
\centering
\begin{tabular}{cc}
\includegraphics[width=0.52\linewidth]{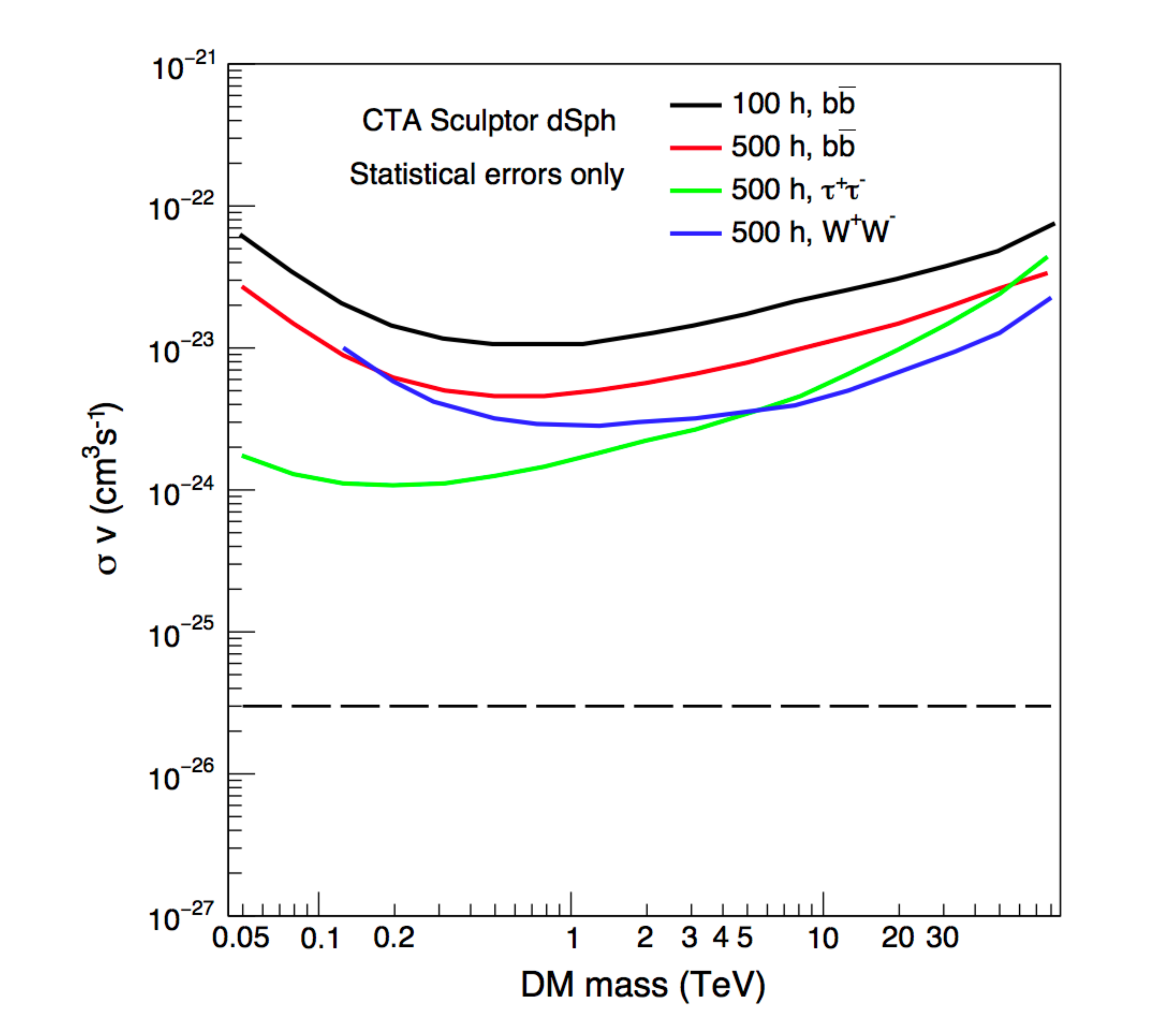}
\includegraphics[width=0.49\linewidth]{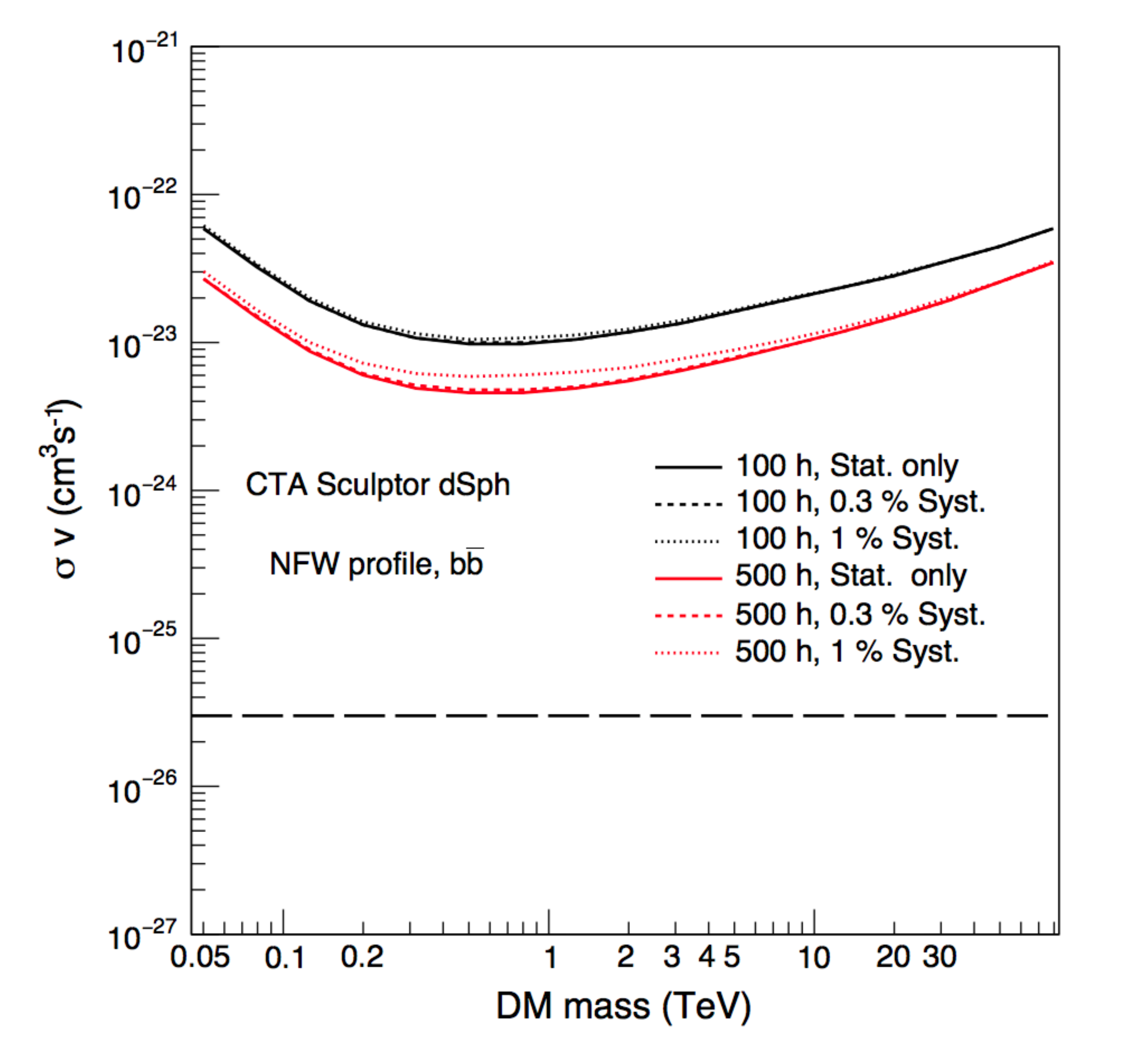}
\end{tabular}
\caption{Left panel: Sensitivity for $<\sigma v>$ from observation of the classical dwarf galaxy Sculptor for the NFW profile and for different annihilation modes as indicated. Right panel: Sensitivity for $<\sigma v>$ for \textit{b\={b}} annihilation modes for different observation times of the same dwarf galaxy. Black line is for 100 h of observation and red line for 500 h. The solid lines are the sensitivities only taking into statistical errors while dashed and dotted curves take into account systematics as indicated in the figure. The dashed horizontal line shows the thermal cross-section in both figures.}
\label{fig:dwarf}
\end{figure*}

The search program selects dwarf galaxy targets on the basis of both their J-factors and relative J-factor uncertainties. Due to the larger available sample of spectroscopically measured stars, the classical dwarf galaxies such as Draco, Ursa Minor, Sculptor, and Fornax have significantly smaller uncertainties on the J-factor than the ultrafaint dwarf galaxies \cite{Fermi_Dwarf}. Our knowledge of the dark matter distribution in the so-called classical dSph satellites of the Milky Way is based on dynamical modelling of their internal stellar kinematics \cite{Bonnivard}.
It is reasonable to expect that on a 5-10 year timescale there will be samples of at least 1000 stars for all of the classical dSphs, with Sculptor and Fornax yielding particularly rich samples of 5K and 10K stars, respectively. These new data will need to be complemented by more advanced modelling in order to constrain their dark matter profiles.  
Studies are underway to estimate the impact that increased data sets and improved modelling can be expected to have on the ability of CTA to constrain the nature of dark matter
The observations of a dwarf spheroidal galaxy will be started in the first year of the Dark Matter Programme. Any hints of dark matter signals or unknown sources would guide the plans for future observations. In the absence of signals, a programme of observation on the most promising dSph would be taken and the observing strategy is to acquire 100 h of observations per year on the best candidate dSph at that time.
An example of the sensitivity which could be obtained by observations of a classical dwarf galaxy is shown in Figure \ref{fig:dwarf}. 
In making this comparison with Figure \ref{fig:halo_fig} it can be seen that the sensitivity is a factor 100 worse for this classical dwarf galaxy; however the effect of systematics is drastically reduced for this small source compared to the extended Galactic Halo, explaining the significant interest in observations of dwarfs. Observations of a classical dSph provide comparable sensitivity to the Galactic centre for the cored dark matter profile and cored profiles in dSphs impact their sensitivity reach by a factor of only a few \cite{CTA17}.

\vskip -8mm
\section{Large Magellanic Cloud}
\vskip -2mm

\begin{figure*}[h!t]
\centering
\begin{tabular}{cc}
\includegraphics[width=0.495\linewidth]{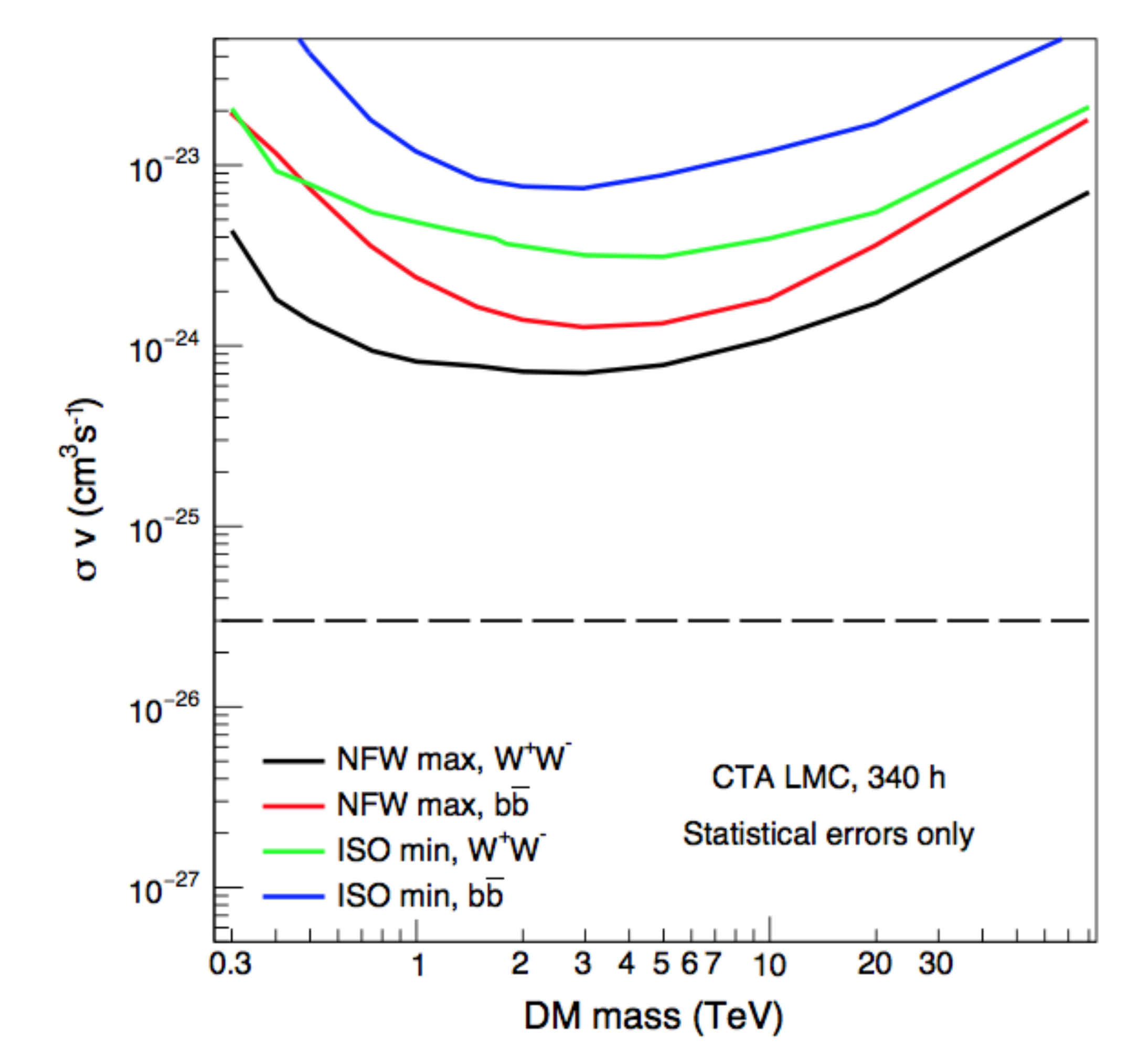}
\includegraphics[width=0.50\linewidth]{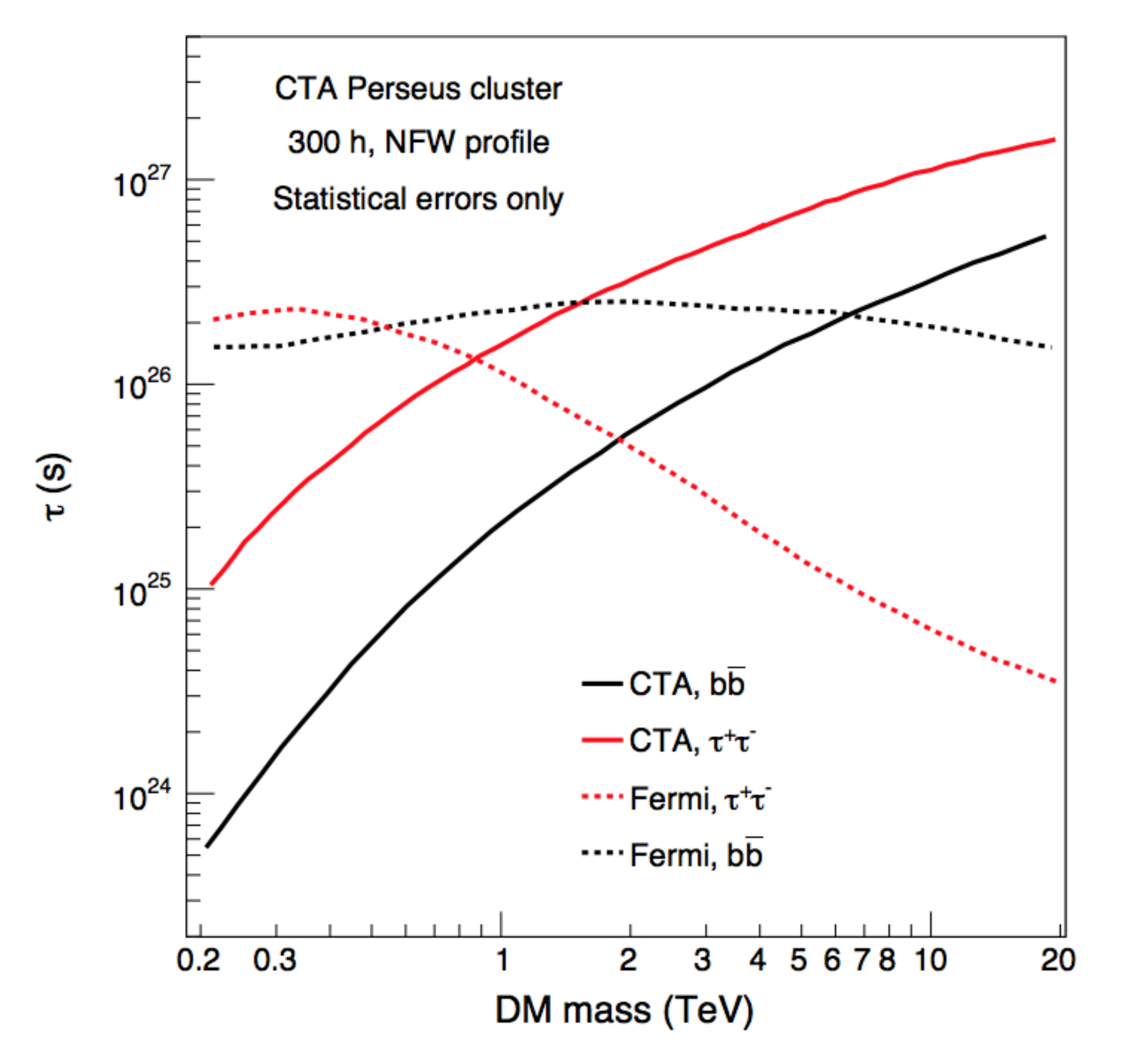}
\end{tabular}
\caption{Left panel: CTA sensitivity on $<\sigma v>$ from observation of the LMC for 340 hours of observation in the \textit{b\={b}} and $W^+W^-$  annihilation channels for both NFW and isothermal (ISO) dark matter profiles, as shown in the legend. The sensitivities are computed with a 200 GeV energy threshold assuming statistical errors only. Right panel: Expected CTA sensitivity to the dark matter decay lifetime $\tau$ for 300 h of observation of the Perseus cluster. We assume a dark matter profile as in \cite{clusters} and adopted the full-likelihood analysis of \cite{method}. The size of the signal integration region ($0.3^\circ$  radius) has been optimised taking into account the expected source extension and the performance for off-axis observations. We assume five off regions. We compare our CTA predictions with the results from the Galactic Halo by Fermi \cite{Halo_Fermi}.}
\label{fig:LMC}
\end{figure*}

The Large Magellanic Cloud (LMC) is a nearby satellite galaxy at high Galactic latitude and it has the shape of a disk seen nearly face-on. At a distance of only $\sim$ 50 kpc, and with a large dark matter mass of $\sim ~ 10^{10} M_\odot$ , the LMC has long been recognized as a potentially favorable target for indirect dark matter searches  \cite{Tasitsiomi}.
For the LMC, observations will be taken with several pointings to cover the full galaxy. A total of 340 h of observations are proposed for both DM and astrophysical motivations. The sensitivity is computed for two benchmark annihilation channels, \textit{b\={b}} and $W^+W^-$ , with the results shown in Figure \ref{fig:LMC} (left). The curves represent the $95 \%$ confidence level upper limits that would be obtained on the dark matter annihilation cross-section as a function of dark matter particle mass in the case that no emission associated with a dark matter spatial template is detected. The minimum energy considered in the analyses is 200 GeV due to the minimum zenith angles allowed for LMC observations. 
The strongest sensitivity is achieved for the NFW profile with the maximum rotation curve and maximum allowed density within uncertainties in the inclination angle of the LMC, while the minimum rotation curve with the isothermal profile yields the weakest sensitivity. In the most optimistic case, the expected sensitivity is about a factor of twenty above the thermal relic cross-section. The difference in the testable annihilation cross-section between the extreme cases is a factor of $\sim 10$. The astrophysical backgrounds in the LMC and their uncertainties differ from those in the Galactic centre making it a complementary dark matter search target that offers competitive sensitivity to annihilation signals.

\section{Clusters of galaxies}

Galaxy clusters are a promising target for decaying dark matter (see, for instance \cite{Cirelli2}). While the signal originating from annihilating dark matter is proportional to the square of the dark matter density, for decaying dark matter the dependence is on the first power. As a consequence, dense dark matter concentrations outshine the astrophysical backgrounds if annihilation is at play, but remain comparatively dim if dark matter is decaying. Decaying dark matter wins instead, generally speaking, when large volumes are considered. Figure \ref{fig:LMC} (right) shows predictions for the case of the Perseus cluster for 300 h of observation. We assume a dark matter profile in the cluster as in \cite{{clusters}} while adopting the full-likelihood analysis of \cite{method}. We considered an integration radius of 0.3$^\circ$. As is clear from the figure, CTA can do much better than Fermi \cite{Halo_Fermi} at the TeV scale.

\section{Conclusion}
The Dark Matter Programme comprises ten years of observations dedicated to dark matter targets. The first three years are devoted to deep observations of the Galactic centre together with observation of the best  dwarf galaxy. These observations  will represent an important achievement and the result, eagerly awaited by the scientific community, is an expected deliverable of CTA.

\section{Acknowledgements}
This work was conducted in the context of the CTA DMEP Group.  \par\noindent
We gratefully acknowledge financial support from the agencies and organizations listed here : http://www.cta-observatory.org/consortium\_acknowledgments

\end{document}